\documentclass[twocolumn,aps,showpacs,preprintnumbers,lettersize,prb]{revtex4}
\usepackage{graphicx}

\begin{document}

\title{Anomalous superconductivity and its competition with antiferromagnetism in doped Mott insulators}
\author{S. S. Kancharla$^{1}$, B. Kyung$^{1}$, David S\'en\'echal$^{1}$, M. Civelli$^{2,3}$, M. Capone$^{4}$, %
G. Kotliar$^{2}$, A.-M.S. Tremblay$^{1}$}
\affiliation{$^{1}$D\'{e}partement de physique and Regroupement qu\'{e}b\'{e}cois sur les
mat\'{e}riaux de pointe, Universit\'{e} de Sherbrooke, Sherbrooke, Qu\'{e}%
bec J1K 2R1, Canada,\\
$^{2}$Department of Physics and Center for Materials Theory, Rutgers
University, Piscataway, NJ, USA,\\
$^{3}$ Theory Group, Insititut Laue Langevin,
Grenoble, France,\\
$^{4}$SMC, CNR-INFM, and Physics Department, University of Roma "La Sapienza", Piazzale A. Moro 2, I-00185, Rome, Italy and ISC-CNR, Via dei
Taurini 19, I-00185, Rome, Italy.}
\date{\today }

\begin{abstract}
Proximity to a Mott insulating phase is likely to be an important
physical ingredient of a theory that aims to describe high-temperature
superconductivity in the cuprates. Quantum cluster methods are well
suited to describe the Mott phase. Hence, as a step towards a quantitative
theory of the competition between antiferromagnetism and $d$-wave superconductivity in the cuprates, we use Cellular Dynamical Mean Field Theory to compute zero temperature properties of the two-dimensional square lattice Hubbard model.  The $d$-wave order parameter is found to scale like the superexchange coupling $J$ for on-site interaction $U$ comparable to or larger than the bandwidth. The order parameter also assumes a dome shape as a function of doping while, by contrast, the gap in the single-particle density of states decreases monotonically with increasing doping. In the presence of a finite second-neighbor hopping $t^{\prime },$ the zero temperature phase diagram displays the electron-hole asymmetric competition between antiferromagnetism and superconductivity that is
observed experimentally in the cuprates. Adding realistic third-neighbor hopping $t^{\prime \prime }$ improves the overall agreement
with the experimental phase diagram. Since band parameters can vary depending on the specific cuprate considered, the sensitivity of the theoretical phase diagram to
band parameters challenges the commonly held assumption that the doping vs $T_{c}/T_{c}^{max}$ phase diagram of the cuprates is universal. The calculated
angle-resolved photoemission spectrum displays the observed
electron-hole asymmetry. The tendency to homogeneous coexistence of the
superconducting and antiferromagnetic order parameters is stronger than
observed in most experiments but consistent with many theoretical results
and with experiments in some layered high-temperature superconductors. Clearly, our calculations reproduce important features of $d$-wave superconductivity in the cuprates that would otherwise be considered anomalous from the point of view of the standard Bardeen-Cooper-Schrieffer approach. At strong coupling, $d$-wave superconductivity and antiferromagnetism appear naturally as two equally important competing instabilities of the normal phase of the same underlying Hamiltonian.
\end{abstract}

\maketitle

\section{Introduction}

Superconductivity in the cuprates and in the layered organics of the BEDT
family is highly anomalous, \textit{i.e.,} it displays a number of properties
that cannot be explained by the Bardeen-Cooper-Schrieffer (BCS) theory
modified for $d$-wave symmetry. For example in the cuprates, superconductivity
emerges upon doping an antiferromagnetic Mott insulator. Moreover, in the so-called underdoped region
near the insulator, experiments show that the gap in the single-particle
density of states decreases upon doping while $T_{c}/T_{c}^{max}$, or the order
parameter, increases, in sharp contrast with expectations from standard BCS
theory \cite{Sutherland:2003}. In the organics, antiferromagnetism and
superconductivity are separated by a first order transition and a Mott
transition separates the corresponding states with no order. For both the organics
and the cuprates, there is much evidence from approximate solutions that the
essential low-energy physics is described by the one-band Hubbard model for
the appropriate lattice, band-structure, interaction and dopings \cite%
{Meinders:1993, Macridin:2005, McKenzieReview:2007}.

Understanding anomalous superconductivity theoretically in a quantitative
manner is still a challenge. An important step towards this goal is to
obtain accurate solutions of the Hubbard model. Despite the apparent
simplicity of the model, it is extremely difficult to solve in the relevant
regime where neither potential ($U$) nor kinetic energy ($8t$) dominate. In
recent years, a number of numerical methods have shed light on this problem.
In this paper, we will describe the results obtained from Cellular Dynamical
Mean-Field Theory \cite{Kotliar:2001} (CDMFT) for $d$-wave superconductivity and its
competition with antiferromagnetism in the cuprates. The corresponding study
in the organics has been published \cite{KyungBEDT:2006, Sahebsara:2006}. The
results will be compared with other quantum cluster methods \cite%
{Maier:2005, LTP:2006, KotliarRMP:2006}, mean-field theories, slave boson, and with variational
approaches.

Our choice of method is motivated by the following considerations. Since
anomalous superconductivity appears near antiferromagnetic Mott insulating
phases, it is important to use approaches that treat these phases correctly.
CDMFT is a generalization of Dynamical Mean-Field Theory (DMFT) \cite%
{Georges:1996,Jarrell:1992}. The latter method describes the Mott insulator-metal
transition of the Hubbard model exactly in the limit of infinite dimensions.
DMFT is, by construction, a local theory that maps the full
interacting many-body lattice problem onto a single-site embedded in a
self-consistent bath. Unfortunately, the local nature of the spatial
correlations inherent to single-site DMFT precludes a description of the
superconducting phase with $d$-wave symmetry observed experimentally in the
cuprates. This limitation has been overcome by several recently developed
cluster extensions of DMFT. In addition to CDMFT, \cite{Kotliar:2001} these include
the Dynamical Cluster Approximation (DCA) \cite{Hettler:1998} and Variational Cluster
Perturbation Theory (VCPT), \cite{Potthoff:2003, Potthoff:2003b}, also known
as Variational Cluster Approximation (VCA). These methods incorporate
short-range correlations as well as some properties of the infinite lattice
in a systematic and causal manner (for reviews see \cite{Maier:2005, LTP:2006, KotliarRMP:2006}). CDMFT maps a lattice problem onto a
finite-size cluster (with different boundary conditions than DCA) and is
able to describe the short-range correlations accurately within the cluster.
A self-consistently determined bath of uncorrelated electrons approximates the effect of the rest of the infinite lattice on
the cluster. Coupling between the embedded cluster and the bath provides a
self-consistent theory that naturally allows for phase transitions and
phases with long range order. In a way, CDMFT makes a compromise between short-range and long-range correlations, the latter being associated with order parameters that appear only in the bath.

Cluster models can be formulated in a general functional framework \cite%
{Potthoff:2003b, KotliarRMP:2006} and they all converge to the same
limit for very large cluster sizes. Nevertheless, different cluster methods
have convergence rates that depend on the physical observable. In
particular, at finite temperatures, CDMFT has been shown to converge
exponentially quickly for local quantities such as the density of states
\cite{Biroli:2005}. The quality of the approximation made by CDMFT has been
tested extensively by comparisons with exact results~\cite{Bolech:2003,Capone:2004}. The
method was found to be accurate in describing the complexities of the
Mott insulator-to-metal transition, indicating that both high as well as low
energy phenomena are well captured. Furthermore, CDMFT has also been used
successfully to elucidate the differences in the evolution of a Mott-Hubbard
insulator into an anomalous correlated metal for the electron- and
hole-doped cases \cite{Parcollet:2004,civelli:2005} as well as to shed light on the
normal state pseudogap phenomenon \cite{KyungPseudogap:2006}.

In the following section, we describe our implementation of CDMFT in more
details. We then present the zero-temperature results and compare with those
of other numerical approaches. It is shown that, in the intermediate
coupling regime $U=8t$, the $d$-wave superconducting phase at zero temperature
is stabilized. In addition, at strong coupling the magnitude of the order
parameter scales with $J=4t^{2}/U.$ We also discuss the effect of
longer-range hopping and of the proximity to the Mott transition. We present
the phase diagram for both hole and electron-doped systems, exhibiting the
competition between $d$-wave superconductivity and antiferromagnetism. The
regions where $d$-wave superconductivity and antiferromagnetism appear are in
semi-quantitative agreement with experiment, except for a region of
coexistence that is not always observed experimentally. It is striking to
observe that as we make the model more realistic by including all the main
band structure parameters $t,t^{\prime }$ and $t\textquotedblright,$ the
coexistence region decreases and the overall phase diagram becomes closer to
experiments. Finally, before we summarize the results, we compute the
single-particle density of states and show that the order parameter
increases with doping in contrast with the single-particle gap that
decreases, in qualitative agreement with experiment. In addition, we find
that the results of Angle Resolved Photoemission Spectroscopy (ARPES) \cite%
{Damascelli:2003} have a natural explanation in our approach. Discussion of our
results in the context of a sample of the existing literature appears at the
end of each subsection.

\section{Model and methods}

We define the Hubbard model Hamiltonian by
\begin{equation}
H=-\sum_{i,j,\sigma }t_{i,j}d_{i,\sigma }^{\dagger }d_{j,\sigma
}+U\sum_{i}n_{i\uparrow }n_{i\downarrow }
\end{equation}%
where $t_{i,j}$ and $U$ correspond to the hopping and the onsite Coulomb
repulsion respectively.
In order to investigate the $d$-wave superconducting phase within CDMFT we
write 
an effective action containing a Weiss dynamical field $\hat{\mathcal{G}}_{0}
$ with both normal (particle-hole) and anomalous (particle-particle)
components that describes the degrees of freedom outside the cluster (the
bath) as a time dependent hopping within the cluster
\begin{equation}
S_{\mathrm{eff}}=\int_{0}^{\beta }d\tau d\tau ^{\prime }\Psi _{d}^{\dagger
}(\tau )\left[ \hat{\mathcal{G}}_{0}^{-1}\right] \Psi _{d}(\tau ^{\prime
})+U\sum_{\mu }\int_{0}^{\beta }d\tau n_{\mu \uparrow }n_{\mu \downarrow }.
\end{equation}%
For the case of a $2\times 2$ plaquette, which we shall consider throughout
this work, the Nambu spinor is defind by
$\Psi _{d}^{\dagger }\equiv (d_{1\uparrow }^{\dagger },\dots ,d_{4\uparrow
}^{\dagger },d_{1\downarrow },\dots ,d_{4\downarrow })$, and the greek letters $%
\mu ,\nu $ label the degrees of freedom within the cluster. Physically, this
action corresponds to a cluster embedded in a self-consistently determined
medium with $d$-wave pairing correlations.

Given the effective action with a starting guess for the Weiss field $\hat{%
\mathcal{G}}_{0}$ we compute the cluster propagator $\widehat{G}_{c}$ by
solving the cluster impurity Hamiltonian that will be described shorty. Then we extract the cluster self
energy from $\hat{\Sigma}_{c}=\hat{\mathcal{G}}_{0}^{-1}-\widehat{G}%
_{c}^{-1} $. Here,
\begin{equation}
\widehat{G}_{c}\left( \tau ,\tau ^{\prime }\right) =\left(
\begin{array}{cc}
\hat{G}_{\uparrow }\left( \tau ,\tau ^{\prime }\right) & \hat{F}\left( \tau
,\tau ^{\prime }\right) \\
\hat{F}^{\dagger }(\tau ,\tau ^{\prime }) & -{\hat{G}}_{\downarrow }\left(
\tau ^{\prime },\tau \right)%
\end{array}%
\right)  \label{nambugreen}
\end{equation}%
is an $8\times 8$ matrix, $G_{\mu \nu ,\sigma }\equiv -\langle Td_{\mu
\sigma }(\tau )d_{\nu \sigma }^{\dagger }(0)\rangle $ and $F_{\mu \nu
}\equiv -\langle Td_{\mu \uparrow }(\tau )d_{\nu \downarrow }(0)\rangle $
are the imaginary-time ordered normal and anomalous Green functions
respectively. Using the self-consistency condition,
\begin{equation}
\hat{\mathcal{G}}_{0}^{-1}(i\omega _{n})=\left[ \frac{N_{c}}{(2\pi )^{2}}%
\int d\tilde{\mathbf{k}}\;\widehat{G}(\widetilde{\mathbf{k}},i\omega _{n})\right]
^{-1}+\hat{\Sigma}_{c}(i\omega _{n})  \label{selfcon}
\end{equation}%
with
\begin{equation}
\widehat{G}(\widetilde{\mathbf{k}},i\omega _{n})=\left[ i\omega _{n}+\mu -\hat{t}(%
\widetilde{\mathbf{k}})-{\hat{\Sigma}_{c}}(i\omega _{n})\right] ^{-1},
\label{Lattice_G}
\end{equation}%
we recompute the Weiss field $\hat{\mathcal{G}}_{0}^{-1}$ and iterate till
convergence. Here $\hat{t}(\widetilde{\mathbf{k}})$ is the Fourier transform of the
superlattice hopping matrix with appropriate sign flip between propagators
for up and down spin and the integral over $\widetilde{\mathbf{k}}$ is performed over
the reduced Brillouin zone of the superlattice.

To solve the cluster impurity problem represented by the effective action
above, we express it in the form of a Hamiltonian $H_{\mathrm{imp}}$ with a
discrete number of bath orbitals coupled to the cluster and use the exact
diagonalization technique (Lanczos method) \cite{Caffarel:1994}
\begin{eqnarray}
H_{\mathrm{imp}} &\equiv &\sum_{\mu \nu \sigma }E_{\mu \nu \sigma }d_{\mu
\sigma }^{\dagger }d_{\nu \sigma }+\sum_{m\sigma }\epsilon _{m\sigma
}^{\alpha }a_{m\sigma }^{\dagger \alpha }a_{m\sigma }^{\alpha }  \nonumber \\
&&+\sum_{m\mu \sigma }V_{m\mu \sigma }^{\alpha }a_{m\sigma }^{\dagger \alpha
}(c_{\mu \sigma }+\mathrm{h.c.})+U\sum_{\mu }n_{\mu \uparrow }n_{\mu
\downarrow }  \nonumber \\
&&+\sum_{\alpha }\Delta ^{\alpha }(a_{1\uparrow }^{\alpha }a_{2\downarrow
}^{\alpha }-a_{2\uparrow }^{\alpha }a_{3\downarrow }^{\alpha }+a_{3\uparrow
}^{\alpha }a_{4\downarrow }^{\alpha }-a_{4\uparrow }^{\alpha }a_{1\downarrow
}^{\alpha }  \nonumber \\
&&+a_{2\uparrow }^{\alpha }a_{1\downarrow }^{\alpha }-a_{3\uparrow }^{\alpha
}a_{2\downarrow }^{\alpha }+a_{4\uparrow }^{\alpha }a_{3\downarrow }^{\alpha
}-a_{1\uparrow }^{\alpha }a_{4\downarrow }^{\alpha }+h.c.).  \nonumber
\end{eqnarray}%
Here $\mu ,\nu =1,...,N_{c}$ label the sites in the cluster and $E_{\mu \nu
\sigma }$ represents the hopping and the chemical potential within the
cluster. The energy levels in the bath are grouped into replicas of the
cluster ($N_{c}=4$) (two replicas in the present case) with the labels $%
m=1,\cdots ,N_{c}$ and $\alpha =1,2$ such that we have $16$ bath energy
levels $\epsilon _{m\sigma }^{\alpha }$ coupled to the cluster via the
bath-cluster hybridization matrix $V_{m\mu \sigma }^{\alpha }$. Using
lattice symmetries we take $V_{m\mu \sigma }^{\alpha }\equiv V^{\alpha
}\delta _{m\mu }$ and $\epsilon _{m\sigma }^{\alpha }\equiv \epsilon
^{\alpha }$. The quantity $\Delta ^{\alpha }$ represents the amplitude of
superconducting correlations in the bath. No static mean-field order
parameter acts directly on the cluster sites \cite{Poilblanc:2002}.

The parameters $\epsilon ^{\alpha }$, $V^{\alpha }$ and $\Delta ^{\alpha }$
are determined by imposing the self-consistency condition in Eq.~\ref%
{selfcon} using a conjugate gradient minimization algorithm with a distance
function
\begin{equation}
d=\sum_{\omega _{n},\mu ,\nu }\left\vert \left( \hat{\mathcal{G}}%
_{0}^{\prime -1}(i\omega _{n})-\hat{\mathcal{G}}_{0}^{-1}(i\omega
_{n})\right) _{\mu \nu }\right\vert ^{2}  \label{distance}
\end{equation}%
that emphasizes the lowest frequencies of the Weiss field by imposing a
sharp cutoff at $\omega _{c}=1.5$. (Energies are given in units of hopping $%
t,$ and we take $\hbar =1$ and $k_{B}=1.)$ The distance function in Eq.(\ref{distance})
is computed on the imaginary frequency axis (effective inverse temperature, $%
\beta =50$) since the Weiss field $\hat{\mathcal{G}}_{0}(i\omega _{n})$ is a
smooth function on that axis.

With the bond superconducting order parameter defined as
\begin{equation}
\psi _{\mu \nu }=\left\langle d_{\mu \uparrow }d_{\nu \downarrow
}\right\rangle
\end{equation}%
we consider $d$-wave singlet pairing ($\psi \equiv \psi _{12}=-\psi
_{23}=\psi _{34}=-\psi _{41}$). The average is taken in the ground state of
the cluster.

To study the competition with antiferromagnetism,
while preserving bipartite symmetry, hybridization $V_{m\mu \sigma
}^{\alpha }\equiv V_{\sigma }^{\alpha }\delta _{m\mu }$ and bath site
energies $\epsilon _{m\sigma }^{\alpha }\equiv \epsilon _{\sigma }^{\alpha }$
become spin dependent. This doubles the number of independent hybridization
and bath parameters. The staggered magnetic order parameter is given by the cluster
wave function average%
\begin{equation}
M_{\mu }=\left\langle d_{\mu \uparrow }^{\dag }d_{\mu \uparrow }-d_{\mu
\downarrow }^{\dag }d_{\mu \downarrow }\right\rangle
\end{equation}%
with $M\equiv M_{1}=-M_{2}=M_{3}=-M_{4}.$

The finite size of the bath in the exact-diagonalization technique is an
additional approximation to the CDMFT scheme. The accuracy of this approximation can be verified by comparing the CDMFT solution for the one-band Hubbard model with the solution from the Bethe ansatz~\cite{Capone:2004, CivelliThesis}. We have also used this comparison in one dimension as a guideline to fix the choice of parameters in the distance function ($\omega _{c}=1.5$ and $%
\beta =50$). These results in one dimension also compare well with those obtained using the
Hirsch-Fye Quantum Monte Carlo algorithm as an impurity solver where the bath is not truncated~\cite%
{KyungQMC:2006}. Further, using
finite-size scaling for these low (but finite) temperature calculations~\cite{KyungQMC:2006}, it was shown that, at intermediate to
strong coupling, a $2\times 2$ cluster in a bath accounts for more than $95\%
$ of the correlation effect of the infinite size cluster in the
single-particle spectrum.

We can also perform an internal consistency check on the effect of the
finite bath on the accuracy of the calculation. With an infinite bath,
convergence insures that the density inside the cluster is identical to the
density computed from the lattice Green function. In practice, we find that
there can be a difference of $\pm 0.02$ between the density estimated from
the lattice and that estimated from the cluster. We will display results as
a function of cluster density since benchmarks with the one-dimensional
Hubbard model show that, with a finite bath and the procedure described
above, one can reproduce quite accurately Bethe ansatz results for $n\left(
\mu \right) $ when the cluster density is used. Nevertheless, we should
adopt a conservative attitude and keep in mind the error estimate mentioned
above.

\section{Anomalous superconductivity and its competition with magnetism}

In this section we first discuss $d$-wave superconductivity by itself.
Competition with magnetism is taken into account in subsequent subsections.
We will conclude with single-particle spectral properties.

\subsection{Scaling as $J$ at strong coupling}

\begin{figure}[tbp]
\includegraphics[angle=0,width=8cm]{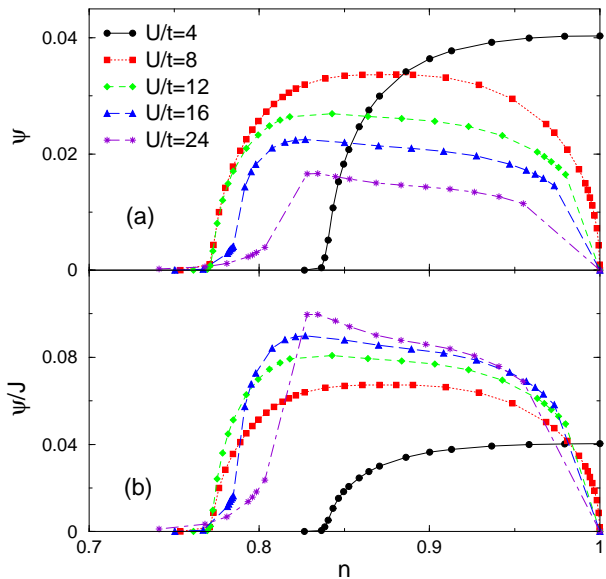}
\caption{(Color online) a) $d$-wave order parameter $\protect\psi $ as a
function of filling $n$ and onsite Coulomb repulsion $U$, $t^{\prime }=0$.
b) Same results but with vertical axis divided by $J=4t^2/U$.}
\label{ordpar}
\end{figure}

The effect of interaction strength on $d$-wave superconductivity is
illustrated in Fig.~\ref{ordpar}\ a, where we show the $d$-wave order
parameter $\psi $ as a function of the density $n$ for $t^{\prime }=0$ and
different values of onsite Coulomb repulsion $U$. No long-range
antiferromagnetism is allowed in this calculation. As $U$ is increased to $%
8t $, the order parameter acquires a very broad maximum around $\delta
=\left\vert 1-n\right\vert =0.15$ (optimal doping). The $d$-wave Weiss field
(not shown), on the other hand, is monotonic as a function of doping with a maximum value close
to half-filling. As the system approaches half-filling from optimal doping
there is a progressive localization due to the increasing proximity of the
Mott insulator and hence, despite the stronger $d$-wave Weiss field in this
region, there is a suppression of the $d$-wave order parameter. This
suppression is not seen for $U=4t$ that is too small for a Mott transition.
However, even in that case, if we were to allow for long-range
antiferromagnetic correlations, the insulating gap at half-filling that
emerges from antiferromagnetic (Slater) correlations should also suppress
$d$-wave superconductivity completely \cite{Kyung:2003, Metzner:2007}.
Clearly, antiferromagnetism near half-filling can destroy superconductivity
but the paramagnetic Mott transition by itself, without antiferromagnetism,
suffices at strong to intermediate coupling. The maximum of the order parameter increases as $U$ increases but
then it decreases as $U$ changes to $12t$, $16t$ and $24t$. This clearly
signals that $d$-wave superconductivity will be strongest at intermediate
coupling and that at strong coupling the $d$-wave order parameter scales as
the superexchange coupling $J=4t^{2}/U$. In Fig.~\ref{ordpar}\ b, the
value of the order parameter is divided by $J$ on the vertical axis to
better exhibit this scaling for $U\geq 12t$. The value of optimal doping is
nearly independent of $U$ in the intermediate to strong coupling regime.

On the overdoped side, $d$-wave superconductivity disappears at a doping
that is comparable with the Dynamical Cluster Approximation (DCA) \cite{Hettler:1998, Maier:2000a} and VCA \cite{Senechal:2005}. Variational calculations
\cite{Paramekanti:2004, Sorella:2002} find $d$-wave superconductivity up to
about $35\%$ doping. The strong-coupling scaling with the superexchange
coupling $J=4t^{2}/U$ as well as the dome like shape of the superconducting
order parameter, which are a hallmark of correlated superconductivity, were
also found in VCA \cite{Senechal:2005}, renormalized mean-field theory \cite%
{Gros:2007}, gauge theories~\cite{Lee:2007} and in early RVB slave boson
studies \cite{Kotliar:1988, Andrei:1988}. In multiorbital models for
fullerenes, a dome behavior is also found as a function of the correlation
strength \cite{Capone:2002,Capone:2004b}. We note that at large doping for $%
U=8t-12t,$ we find in Fig.~\ref{ordpar} that the fall of the $d$-wave order
parameter occurs with constant negative curvature as a function of doping,
as seems to occur experimentally. This contrasts with weak coupling studies
\cite{Kyung:2003, Metzner:2007} where this does not occur.

Recent experiments \cite{Keren:2006} in four families (different $x$) of
high-temperature superconductors (Ca$_{x}$La$_{1-x}$)(Ba$_{1.75-x}$La$%
_{0.25+x}$)Cu$_{3}$O$_{y}$ suggest that scaling with $J$ has been observed
experimentally in the cuprates.

\subsection{Effect of second-neighbor hopping $t^{\prime }/t$ without
competition with antiferromagnetism}

\begin{figure}[tbp]
\includegraphics[width=6cm,angle=0]{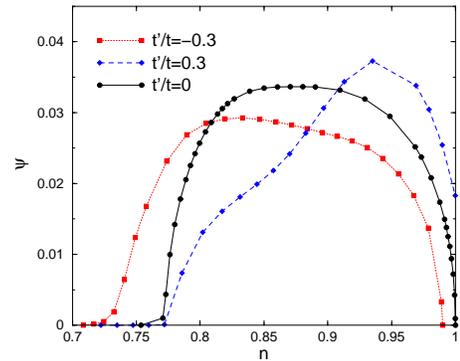}
\caption{$d$-wave order parameter for $U=8t$ as a function of filling $n$ for
various values of $t^\prime /t$ (frustration).}
\label{frust}
\end{figure}

The effect of next-nearest-neighbor hopping $t^{\prime }/t$ is illustrated
in Fig.~\ref{frust} where we plot the $d$-wave order parameter $\psi $ as a
function of doping for various values of the next-nearest neighbor
(diagonal) hopping $t^{\prime }$. For clarity, we represent both the hole
and electron-doped cases on the same plot by performing a particle-hole
transformation and considering $t^{\prime }=+0.3t$ in the electron-doped
case. The maximum of $\psi $ grows with increasing $t^{\prime }$.

This trend with $t^{\prime }$ was found in DCA \cite{Maier:2000a}, in
recent variational studies \cite{Sorella:2007} and in two-leg ladder studies
of the $t-t^{\prime }-J$ model \cite{White:1999}. However, the increase of $%
\psi $ with $t^{\prime }$ does not seem consistent with the empirical
correlation found in Ref. \onlinecite{Andersen:2001} between a larger optimal $T_{c}$
and a smaller value of the $t^{\prime }$ obtained from band structure
calculations. As found in an earlier VCA study \cite{Senechal:2005} the
maximum $\psi $ is larger for electron doping than for hole doping when calculations are done on
$2\times 2$ clusters. However, the situation reverses for larger clusters.
This suggests that finite-size effects may be influencing our conclusions on
the effect of $t^{\prime }$. Superconductivity is known to be strongest on
the hole- as opposed to the electron-doped compounds (Using particle-hole
transformation, the latter correspond to positive $t^{\prime }$ on Fig.\ref%
{frust}). Nevertheless, we will see in the following section that including
the competition with antiferromagnetism restores the experimentally observed
electron-hole asymmetry.

\subsection{Phase diagram and competition with magnetism}

\begin{figure}[tbp]
\includegraphics[angle=0,width=7cm]{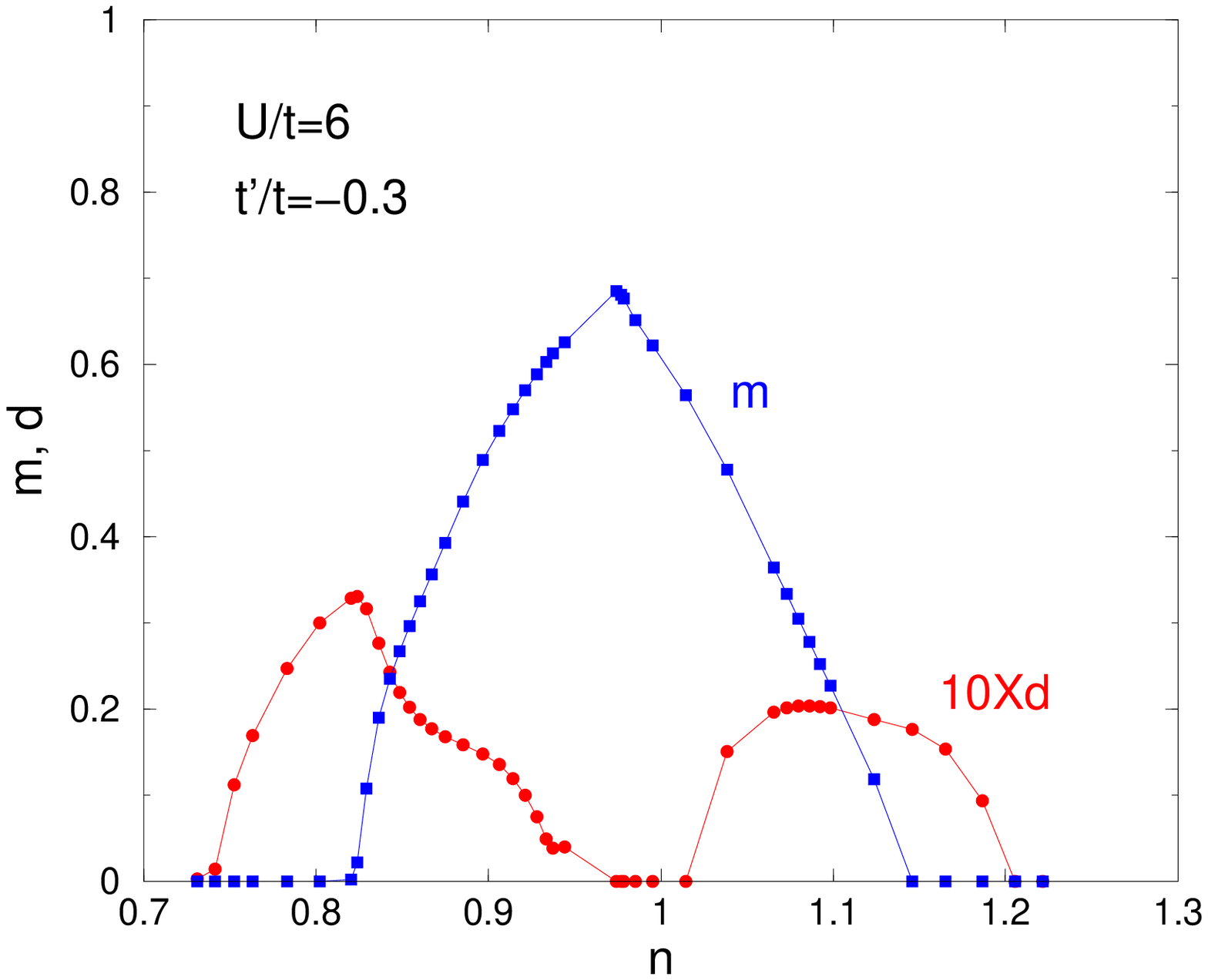} %
\includegraphics[angle=0,width=7cm]{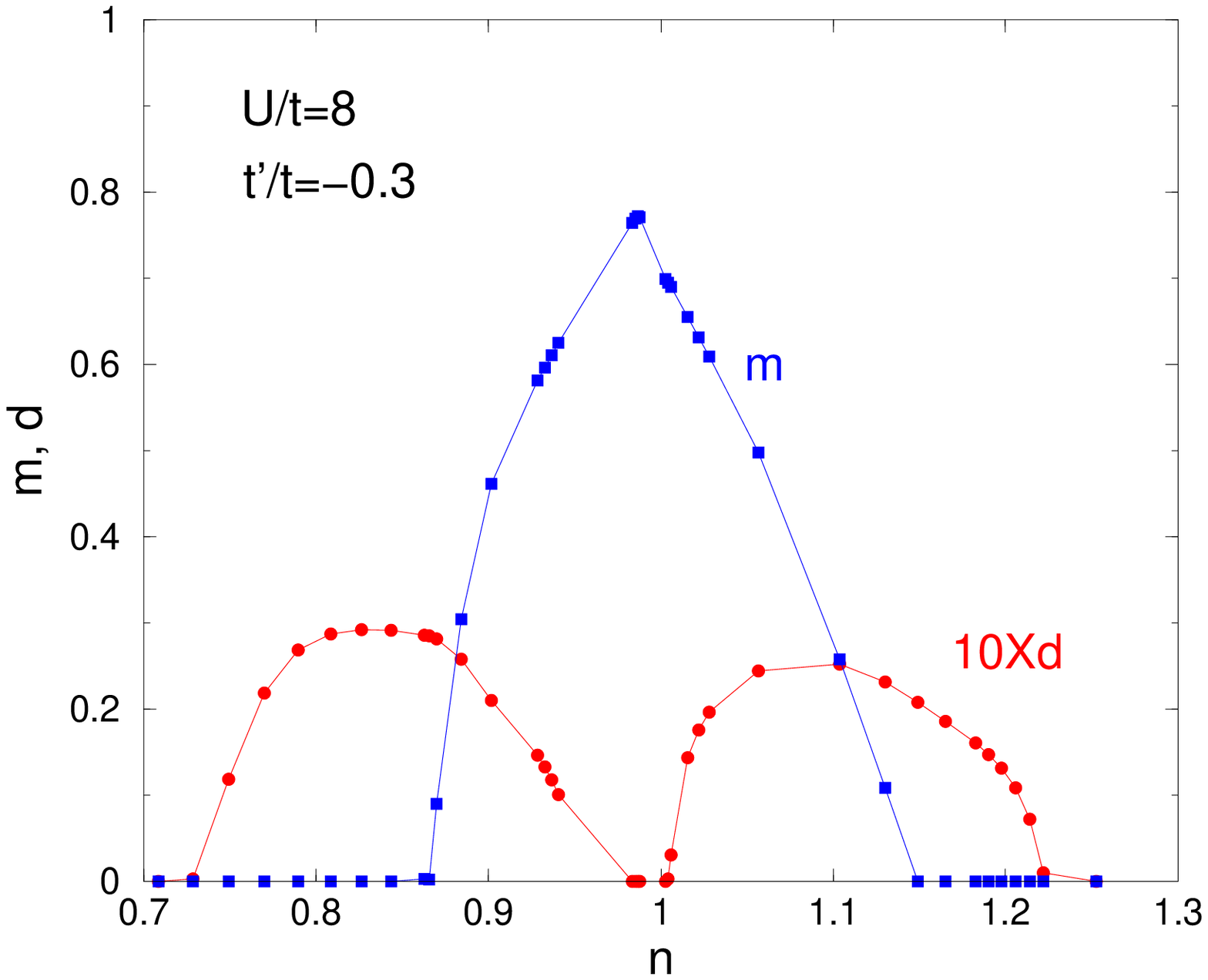} %
\includegraphics[angle=0,width=7cm]{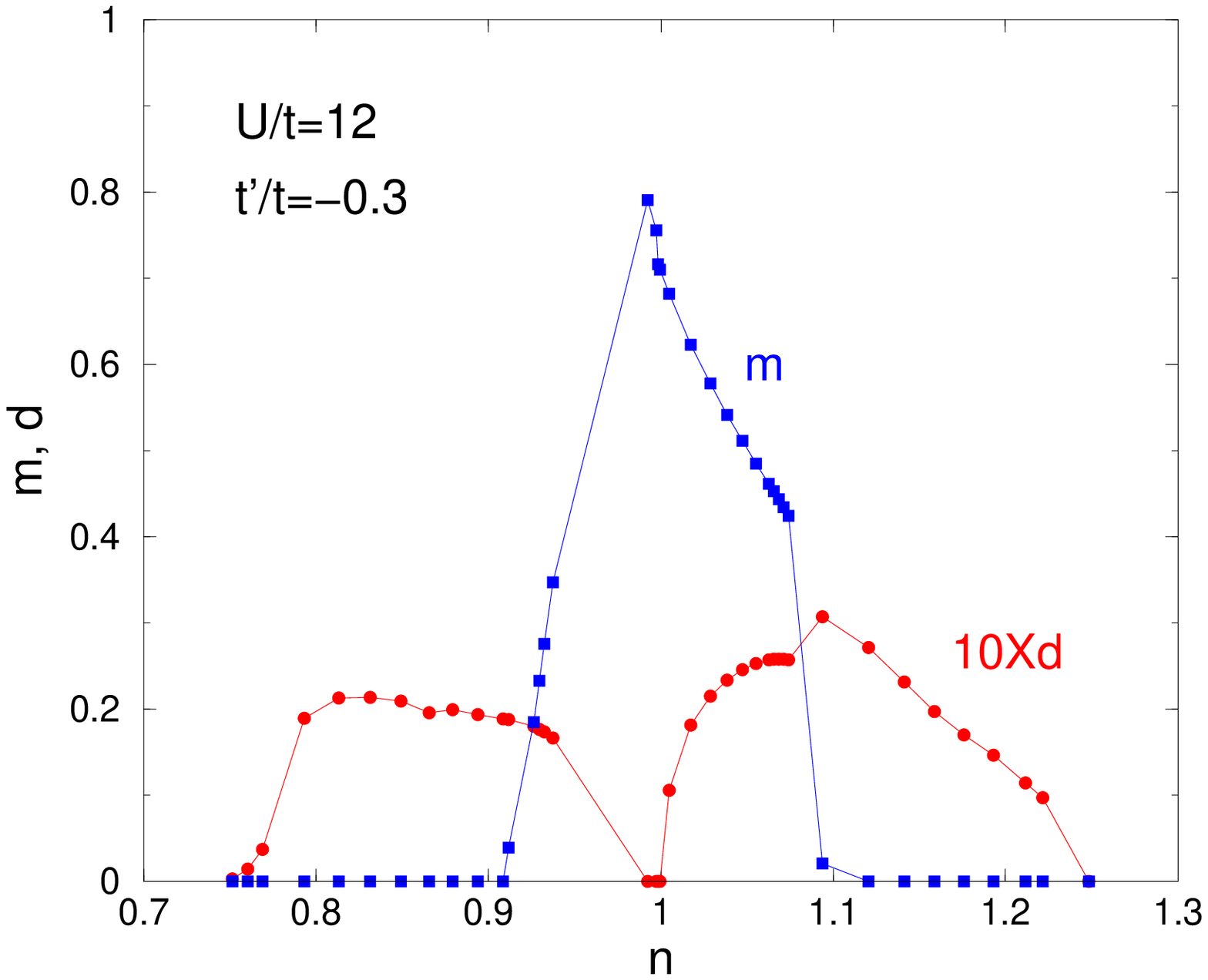}
\caption{(Color online) Superconducting (red circles) and antiferromagnetic (blue squares)
order parameters for $t^{\prime }=-0.3t$ and, from top to bottom, the three
values $U=6t,$ $U=8t,$ and $U=12t$. The amplitude of the superconducting
order parameter is multiplied by a factor of 10 to be on a scale comparable
with the antiferromagnetic one.}
\label{Diag_phase}
\end{figure}

In the previous sections, we have constrained the bath of the associated
impurity problem to allow only normal state and $d$-wave superconducting
solutions. In this section, we start from initial bath parameters that allow
for both $d$-wave and antiferromagnetic symmetry breaking. Fig. \ref%
{Diag_phase} displays the results for the zero-temperature phase diagram for
$t^{\prime }=-0.3t$ and three values of the interaction strength $U=6t,$ $%
U=8t$ and $U=12t.$ The plots are as a function of electron density, so that
the system is hole-doped for $n<1$ and electron-doped for $n>1$. The order
parameter for $d$-wave superconductivity is multiplied by a factor of $10$ so
that it is on the same scale as the antiferromagnetic order parameter. We
have checked that in the coexistence region, even if we start the iterations
with a fully converged antiferromagnetic solution, the final solution is the
same as the one exhibited in Fig.\ \ref{Diag_phase}.

From Fig. \ref{Diag_phase}, we observe that antiferromagnetism occurs over a
narrower range of dopings as $U$ increases since $J$ then decreases (the
trend would be opposite at weak coupling). There is homogeneous coexistence
of antiferromagnetism and $d$-wave superconductivity near half-filling in all
cases. That phase can be called a superconducting antiferromagnet or an antiferromagnetic superconductor. $D$-wave superconductivity
exists by itself at large electron or hole doping. The transition from a
homogeneous coexistence phase to pure $d$-wave superconductivity is second
order, except for large values of $U=12t$ on the electron-doped side.
Qualitatively, it seems that $U=8t$ gives a better agreement with the
experimental phase diagram of the cuprates since, in that case,
superconductivity appears alone over a broader range of dopings on the hole
than on the electron-doped side. Also, the value of the maximum $d$-wave order
parameter is larger on the hole- than on the electron-doped side, showing
that competition with antiferromagnetism can reverse the trend observed as a
function of $t^{\prime }$ in Fig.\ref{frust}. Choosing a value of $U$ on the
electron-doped side \cite{Senechal:2004,Dare:2004} that is smaller than the
value of \thinspace $U$ for the hole-doped side would also help in making
the tendency for $d$-wave superconductivity smaller on the electron-doped side,
as seems to be observed experimentally. The asymmetry in the maximum value
of the $d$-wave order parameter for hole and electron doping is also observed
for $U=6t$, but in that case the antiferromagnetism occurs over a doping
range that is unreasonably large compared with experiment. For $U=8t$,
optimal doping occurs around $15\%$ in the hole-doped case. $U=8t$ is also consistent with the value necessary to explain details of the spin wave spectrum obtained by neutron measurements at half-filling~\cite{Coldea:2001,Delannoy:2007}. All these
qualitative trends agree with the experimental phase diagram except for the
following: Antiferromagnetism extends over a broader range of dopings on the
hole-doped side than observed experimentally (see however the next section)
and there is a strong tendency for homogeneous coexistence of the two order
parameters, even though the $d$-wave order parameter is suppressed by the
presence of antiferromagnetic order. While the suppression of $d$-wave
superconductivity by antiferromagnetism is appreciable, the reverse effect
is almost negligible \cite{LTP_AFMvsDSc}. This is also observed in
mean-field studies \cite{Kyung:2000}.

Homogeneous coexistence of antiferromagnetism and $d$-wave superconductivity is not generic in the cuprates. Nevertheless it
has been observed recently in ordered layered compounds \cite{Mukuda:2006}
and in the electron-doped cuprate PrCeCuO \cite{Dai:2005}. Although such
coexistence is not observed in compounds like YBCO, it appears in La$_{2}$CuO%
$_{4.11}$ at zero field and in La$_{1.9}$Sr$_{0.1}$CuO$_{4}$ under applied
magnetic field \cite{Birgeneau:2006}. The agreement of the latter
field-dependent experiments \cite{Khaykovich:2002} with the theoretical
predictions \cite{Demler:2002} reveals the proximity of single-phase $d$-wave
superconductivity with homogeneous coexistence of antiferromagnetism and
$d$-wave superconductivity. Other LSCO compounds \cite{Lake:2001, Mesot:2007}
reveal the proximity of antiferromagnetism and $d$-wave superconductivity
through the application of a magnetic field. Muon spin rotation studies as a function of doping, temperature and field show that zero-field coexistence of superconductivity with short-range static magnetism~\cite{Sonier:2007} is generic in the underdoped regime. Since competition between antiferromagnetism and $d$-wave superconductivity seems to be quite sensitive to disorder~\cite{Dagotto:2005}, the homogeneous coexistence between the two phases that we find here may be reconcilable with experiments.


It is instructive to compare our theoretical results with other approaches
that include competition of $d$-wave superconductivity with
antiferromagnetism. Early variational calculations~\cite{Giamarchi:1990} at $t^{\prime }=0,$ $%
U=10$ and with the $t-J$ model \cite{Himeda:1999}
showed a strong tendency for antiferromagnetism and superconductivity to
coexist homogeneously, even though there is a region where $d$-wave
superconductivity exists by itself. The most recent variational calculations
\cite{Sorella:2007,Randeria:2007} show the same trend. Weak coupling
calculations with functional renormalization coupled to renormalized
mean-field theory at $U=2.5$ and $t^{\prime }=-0.15$ show also a coexistence
region \cite{Metzner:2007}. A recent review \cite{Gros:2007} of variational
approaches and renormalized mean-field theory results \cite{Rice:2007} shows
that a coexistence region is often obtained in these approaches, as it was
in early slave-boson theories \cite{Andrei:1988}.

An early version of cluster DMFT \cite{Lichtenstein:2000} using $t^{\prime }=-0.15t$ and $%
U=4.8t$ found that $d$-wave superconductivity and antiferromagnetism coexist
over most of the doping range $1-n<0.3$. Recent CDMFT calculations \cite%
{Capone:2006} at $t^{\prime }=0$ find that an homogeneous coexistence region
appears at weak coupling but is replaced by a first order transition between
both phases at strong coupling. Phase separation as a function of doping
takes place in strong coupling but not in weak coupling. Here, at finite $%
t^{\prime }$, we did find a first-order transition at large $U$ on the
electron-doped side but phase coexistence persists at low doping.

In VCA calculations that include cluster chemical potential as a variational
parameter \cite{Aichhorn:2006, Aichhorn:2006b} to insure thermodynamic
consistency, a coexistence phase with continuous transition to $d$-wave
superconductivity is found for $t^{\prime }=-0.3$ at weak coupling, but at
strong coupling the transition between homogeneous coexistence and pure
$d$-wave superconductivity is first order \cite{Aichhorn:2007}, as we have
found here for $U=12t$ on the electron-doped side. In early VCA studies \cite%
{Aichhorn:2005b}, $U=8,$ $t^{\prime }=-0.3,$ it was noticed that the energy
scale for that first order transition is larger on the hole than on the
electron doped side. In VCA calculations that do not include chemical
potential as a variational parameter, zero-temperature finite-size studies
find antiferromagnetism and $d$-wave superconductivity over doping ranges that
agree semi-quantitatively with experiment for both hole and electron doped
systems \cite{Senechal:2005}, using $U=8t$ and taking second nearest
neighbor hopping parameter $t^{\prime }=-0.3t$ and third-nearest neighbor
hopping $t^{\prime \prime }=0.2t$ from band structure. As in our case, there
is a coexistence region on the electron-doped side, but on the hole-doped
side there is still some non-monotonic size dependence that suggests that
inhomogeneous phases may be more stable.

Finally, in DCA, system sizes up to $32$ sites have been studied for $U=4t$,
$t^{\prime }=0.$ At $10\%$ doping, a superconducting transition is found~\cite{maier_d:2005}
with critical temperature $T_{c}=0.023$. The question of
ground state coexistence was not studied since these are finite temperature
studies. However, it was established at intermediate coupling that the
frequency and momentum dependence of the pairing channel comes from the
antiferromagnetic fluctuations \cite{Maier:2007a,Maier:2007b,Maier:2007c}. Within DCA, there is a
tendency for phase separation in the normal state on the electron-doped side
of the phase diagram \cite{Macridin:2006}. The same tendency is observed at
weak coupling \cite{Roy:2007} using the Two-Particle Self-Consistent (TPSC)
approach \cite{Vilk:1997}.

\subsection{Additional effects of the band structure on the phase diagram}

\begin{figure}[tbp]
\includegraphics[angle=0,width=7cm]{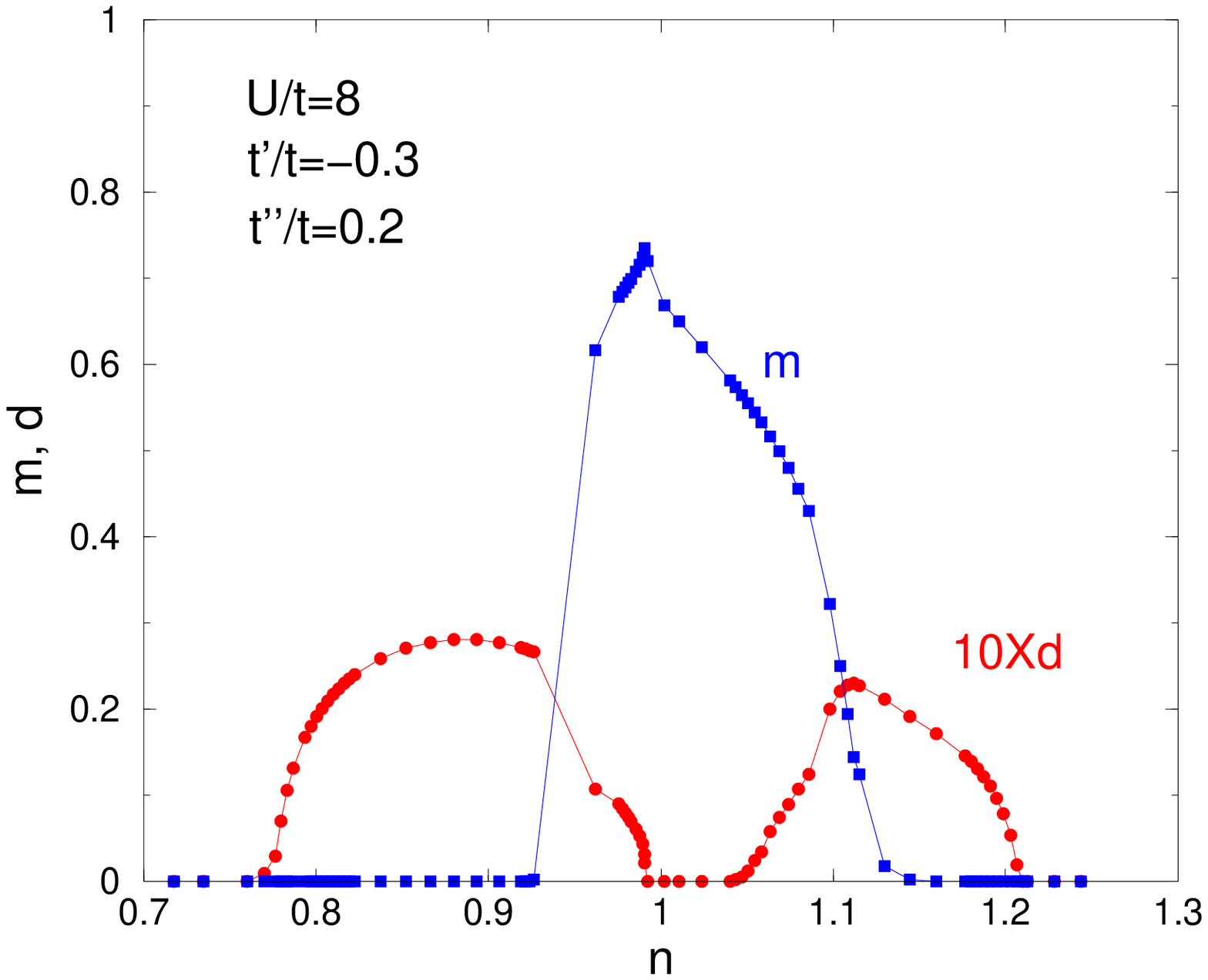} %
\includegraphics[angle=0,width=7cm]{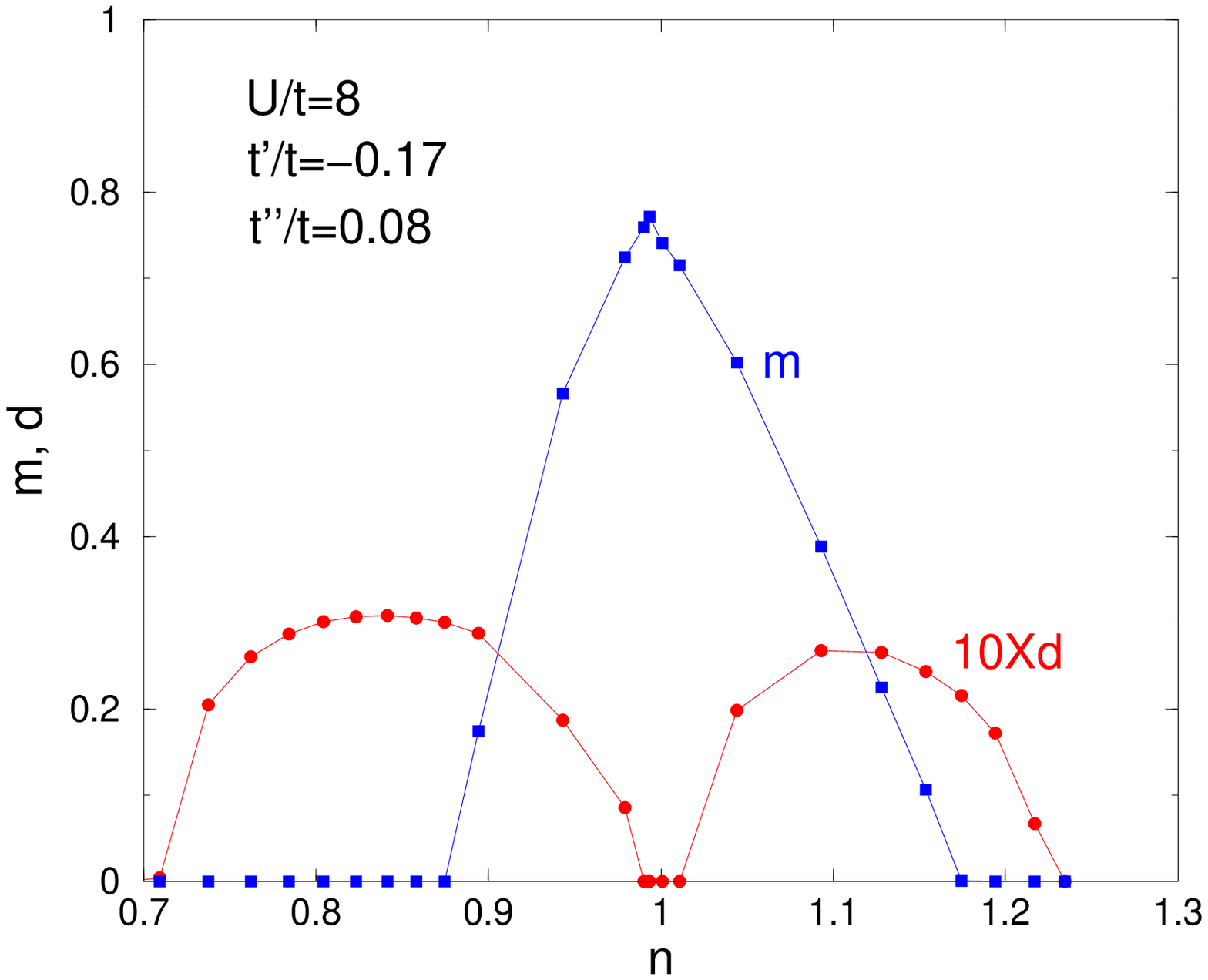}
\caption{(Color online) Superconducting (red circles) and antiferromagnetic (blue squares)
order parameters for $U=8t$ and various values of the band parameters. On
the top, parameters $t^{\prime }=-0.3t,$ $t^{\prime \prime
}=0.2t$ appropriate for YBCO. On the bottom, $t^{\prime }=-0.17t,$ $t^{\prime \prime }=0.08t$ as
in LSCO. The amplitude of the superconducting order parameter is multiplied
by a factor of 10 to be on a comparable scale with the antiferromagnetic
one. }
\label{PhaseDiag_t'-0.3_t''0.2.eps}
\end{figure}

The one-band Hubbard model is an effective model of the cuprates, so
parameters can vary with doping \cite{Senechal:2004,Dare:2004}. Given our
numerical uncertainties, it would be inappropriate at this point to
fine-tune our parameters to try to find a best fit of the experimental phase
diagram of the high-temperature superconductors. Nevertheless, we point out
in this section that when more realistic band parameters are used for the
calculation, the qualitative changes of the resulting phase diagram go in
the right direction to improve agreement with experiment. We also comment on
the implications of sensitivity to band parameters.

The most common parameters used for the one-band model of YBCO for example
\cite{Andersen:2001,Markiewicz:2005} are $t^{\prime }=-0.3t,$ $t^{\prime \prime
}=0.2t$, where $t^{\prime \prime }$ stands for third nearest neighbor hopping. It
has been known for a decade \cite{Leung:1997} that for the half-filled
parent compound Sr$_{2}$CuO$_{2}$Cl$_{2}$, for example, these values
describe well ARPES experiments. Adding third-nearest neighbor hopping $%
t^{\prime \prime }=0.2t$ to the calculation of the previous section leads to
the phase diagram illustrated on the upper part of Fig. \ref%
{PhaseDiag_t'-0.3_t''0.2.eps}. Clearly the phase diagram becomes
qualitatively closer to that of the cuprates since the antiferromagnetic
phase has a stronger electron-hole asymmetry and the tendency for coexistence
between antiferromagnetism and $d$-wave superconductivity is reduced. These
two results were found first in recent VCA calculations \cite{Guillot:2007}.
Note also that the phase boundaries for antiferromagnetism are extremely
close to those obtained from VCA for the same parameters \cite{Senechal:2005}%
. In our CDMFT calculations, the effect of $t^{\prime \prime }$ enters only
in the self-consistency condition Eq.(\ref{selfcon}) through the parameter $%
\hat{t}(\widetilde{\mathbf{k}})$ of the superlattice Green function Eq.(\ref%
{Lattice_G}). In VCA the clusters are larger so that $t^{\prime \prime }$
can also enter in the cluster calculation, not only in the self-consistency
condition. The excellent agreement between both methods for the boundaries
of the antiferromagnetic phases suggests that the main effect of $t^{\prime
\prime }$ is taken into account in our CDMFT calculations.

On the lower part of Fig. \ref{PhaseDiag_t'-0.3_t''0.2.eps}, one finds the
phase diagram corresponding to parameters typical of La$_{2-x}$Sr$_{x}$CuO$%
_{4}$ (LSCO) \cite{Andersen:2001}, namely $t^{\prime }=-0.17t$ and $t^{\prime
\prime }=0.08t.$ There are quantitative changes compared with the top figure
appropriate for YBa$_{2}$Cu$_{3}$O$_{7-x}$ (YBCO). In particular, on the
hole-doped side of the phase diagram the $d$-wave phase extends to larger
dopings for the above LSCO parameters than those for YBCO. Although there is
no qualitative change, this quantitative sensitivity of the phase diagram to
band parameters suggests that the universality of the $T_{c}\left( \delta
\right)/T_{c}^{max}$ relation~\cite{Presland:1991,Obertelli:1992}, usually assumed for
hole-doped ($\delta$) high-temperature superconductors, may not be completely accurate, unless it
signals that band parameters are in fact quite close, as has been suggested in
some recent studies \cite{Delannoy:2007}. The possible lack of universality of the $T_{c}\left( \delta \right)/T_{c}^{max}$ relation has been noticed before~\cite{Markiewicz:2005}. In experiments, it has been pointed out a while ago~\cite{Ando:2000} that single layer Bi$_{2}$Sr$_{2-z}$La$_{z}$Cu$_{2}$O$_{6+\delta}$ (BSLCO) deviates from the universal $T_{c}\left( \delta \right)/T_{c}^{max}$ curve. The electronic structure of that compound differs~\cite{Kondo:2005} from that of LSCO, the compound used to determine the universal $T_{c}\left( \delta \right)/T_{c}^{max}$ curve~\cite{Presland:1991}. Hence, band structure could be one of the factors explaining the deviations of single layer BSLCO. There is also a ``universal'' $T_{c}(S(290 K))/T_{c}^{max}$ determined from thermopower $S(290K)$ at $290 K$ instead of doping~\cite{Obertelli:1992}. However, thermopower has been experimentally linked to band structure~\cite{Okada:2005}.

\subsection{Gap in total density of states and ARPES spectrum}

Having established that the $d$-wave order parameter $\psi $ decreases towards
zero as we approach half-filling, we now show that the gap $%
\Delta $ computed from the single-particle density of states increases
monotonically towards half-filling $n=1$ in the hole-doped case. All the
calculations of this section are for $t^{\prime }=-0.3t$ and $U=8t.$ The
local density of states can be obtained from the cluster or from the lattice
Green function. The results are identical in the limit of infinite bath size. The
single-particle gap is extracted from the local density of states, that we
obtain from the lattice Green function that will be used later to obtain
the single-particle spectral weight $A\left(\mathbf{k},\omega =0\right) $. The lattice Green function is
computed by periodizing the superlattice Green function $G^{\mu \nu }(%
\widetilde{\mathbf{k}},\omega )$ in Eq.(\ref{Lattice_G}) over the entire Brillouin
zone of the original infinite lattice as is done in CPT \cite{Senechal:2004}. This proceeds as follows. In reciprocal space, any wave vector $\mathbf{k}$ in the
Brillouin zone may be written as $\mathbf{k}=\widetilde{\mathbf{k}}+\mathbf{K}$ where both $\mathbf{k}$ and $%
\widetilde{\mathbf{k}}$ are continuous in the infinite size limit. However $%
\widetilde{\mathbf{k}}$ is defined only in the reduced Brillouin zone that
corresponds to the superlattice. On the other hand, $\mathbf{K}$ is discrete and
denotes reciprocal lattice vectors of the superlattice. Then the lattice
Green function is defined by
\begin{equation}
G_{\mathrm{latt}}(\mathbf{k},\omega )=\sum_{\mu ,\nu }e^{-i\mathbf{k}\cdot (r_\mu -r_\nu )}G^{\mu
\nu }(\mathbf{k},\omega ),  \label{transf}
\end{equation}%
where $G^{\mu \nu }(\mathbf{k},\omega )=G^{\mu \nu }(\widetilde{\mathbf{k}},\omega ).$
Integration over $\mathbf{k}$ yields the local density of states. Different
periodization procedures can be used \cite{KotliarRMP:2006, LTP:2006}. %

\begin{figure}[tbp]
\includegraphics[angle=0,width=7cm]{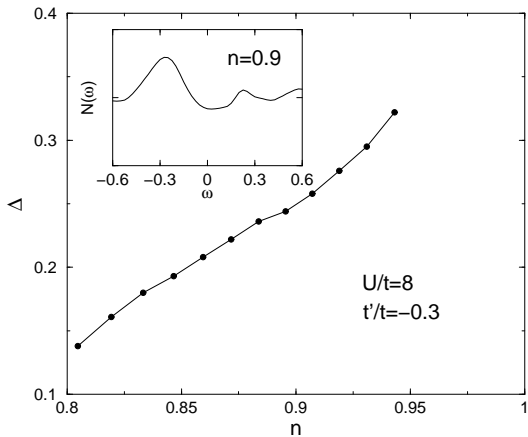}
\caption{The gap as a function of filling, for $U=8t$, $t^{\prime }=-0.3t$.
The gap is defined as half the distance between the two peaks on either side of
$\protect\omega =0$, as they appear, for example, in the inset.}
\label{gap}
\end{figure}

The single-particle gap is defined as half the distance between the peaks
that surround the minimum near $\omega =0.$ The inset of Fig.~\ref{gap} shows the local density
of states in the superconducting state at $n=0.9.$ Due to the finite value
of the broadening parameter $(0.1t)$, some spectral weight fills the $d$-wave
gap, which vanishes linearly near $\omega =0$. We do not focus on this $%
\omega \rightarrow 0$ energy scale. In Fig.~\ref{gap} we exhibit the
monotonic increase of the single-particle gap as filling increases toward $n=1$
(unlike the order parameter decrease) in accordance with, for example,
thermal conductivity \cite{Sutherland:2003} and tunneling experiments \cite%
{Renner:1998, Timusk:1999} for the cuprates. This is a radical departure
from a conventional superconductor where the single-particle gap $\Delta $
and $\psi $ are proportional to each other. Experimentally in the hole-doped cuprates
\cite{Sutherland:2003}, the magnitude of the gap at optimal doping is estimated to
be around $50meV$. Here we obtain $0.2t,$ which is in good agreement with
experimental estimates if we take a reasonable value $t=250meV$. CDMFT
should be reliable for the value of this gap since it is extracted from a
local quantity. 

\begin{figure}[tbp]
\resizebox{6.0cm}{!}{%
\includegraphics{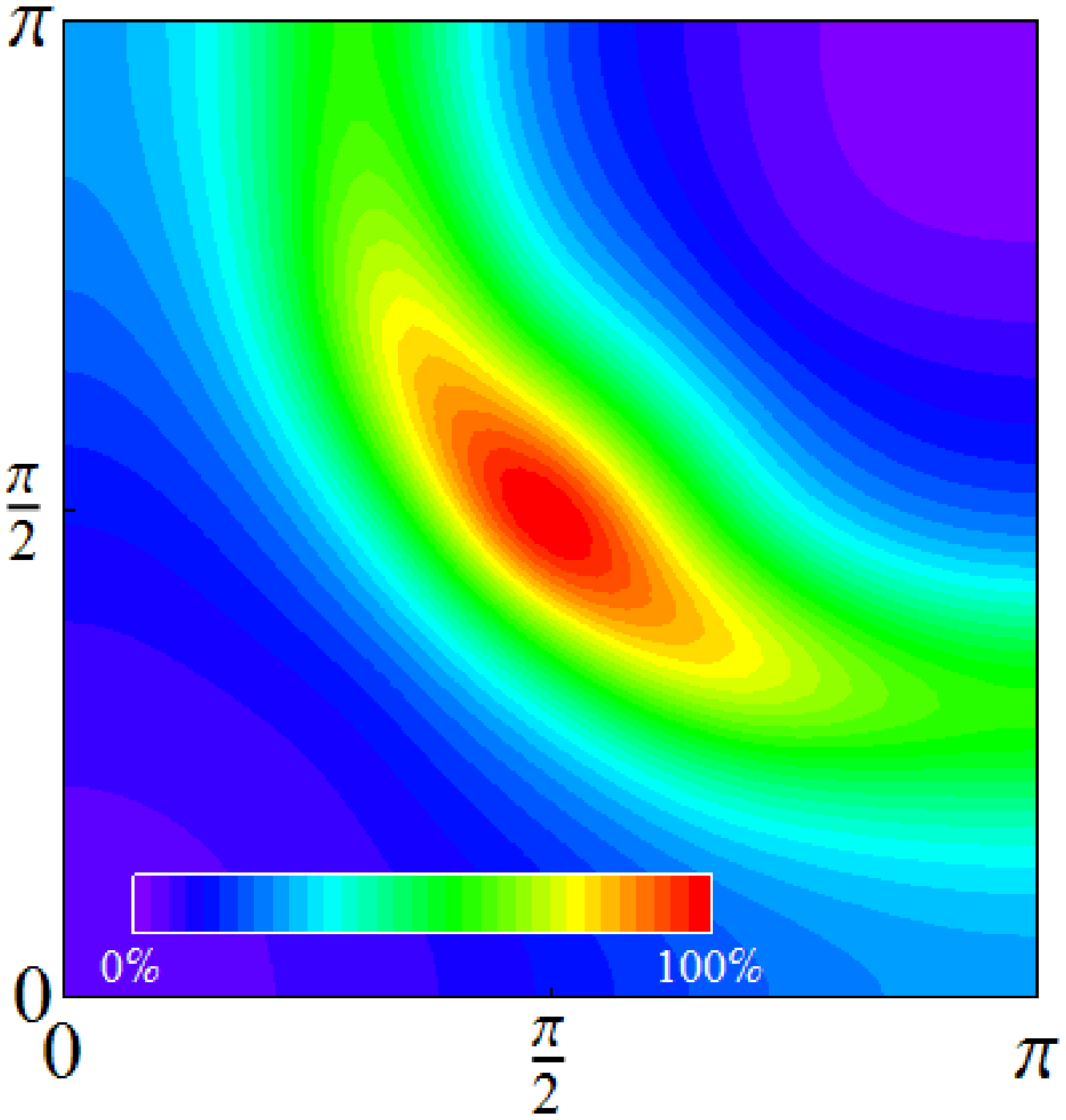}} %
\includegraphics[angle=-90,width=7cm]{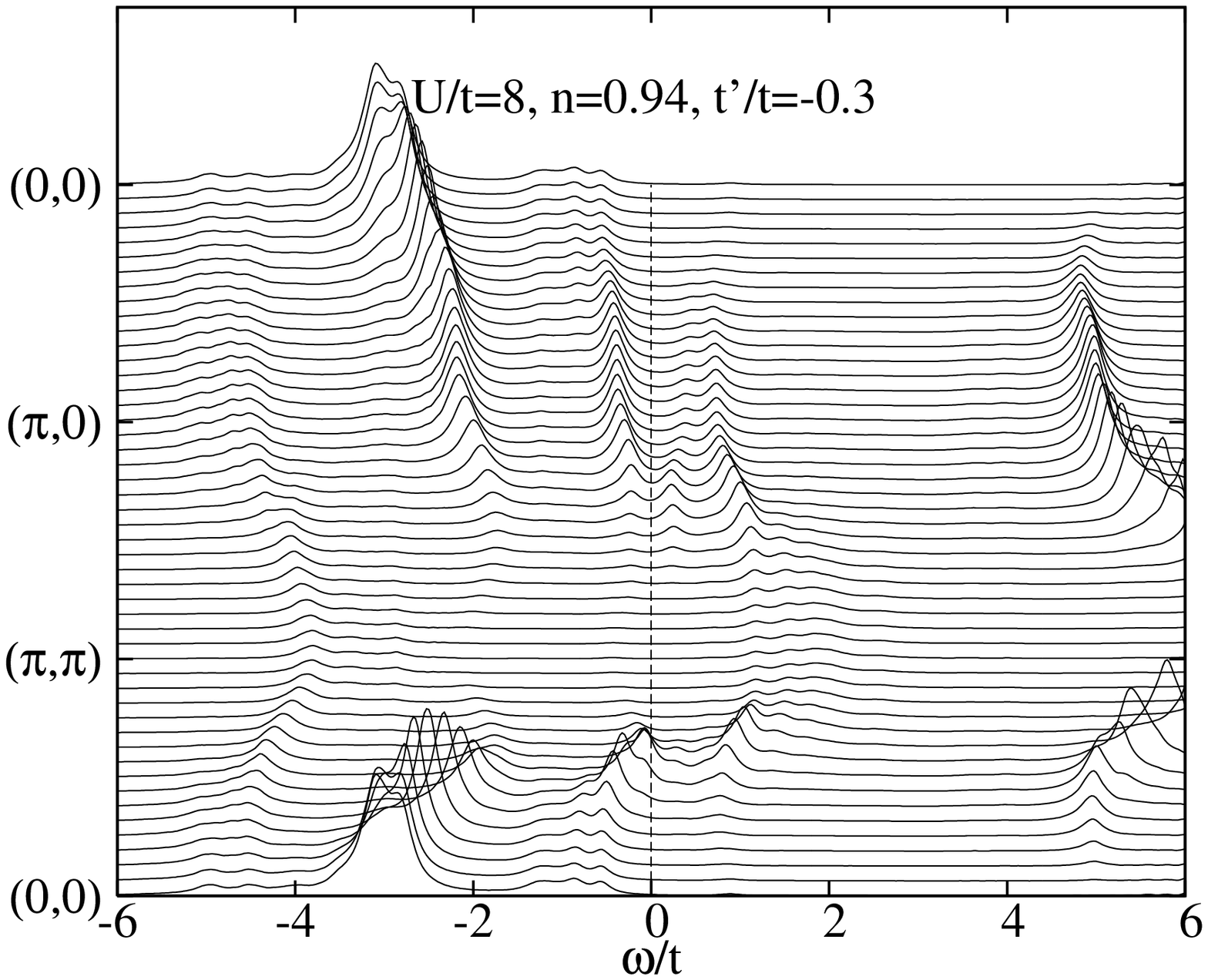}
\caption{(Color online) Hole-doped $d$-wave superconducting phase for $n=0.94,$
$U=8t,$ $t^{\prime }=-0.3.$ Upper figure, color contour plot of $A\left( \mathbf{k},\protect\omega %
=0^{+}\right) $ in first quadrant of the Brillouin zone. Energy resolution
is $\protect\eta =0.1t$ for both figures. Lower figure, $A\left( \mathbf{k},%
\protect\omega \right) $ for various values of wave vector along
high-symmetry directions. The relative scale in the contour plot can be made
quantitative by comparison with the lower figure. }
\label{spectra}
\end{figure}

\begin{figure}[tbp]
\resizebox{6.0cm}{!}{%
\includegraphics{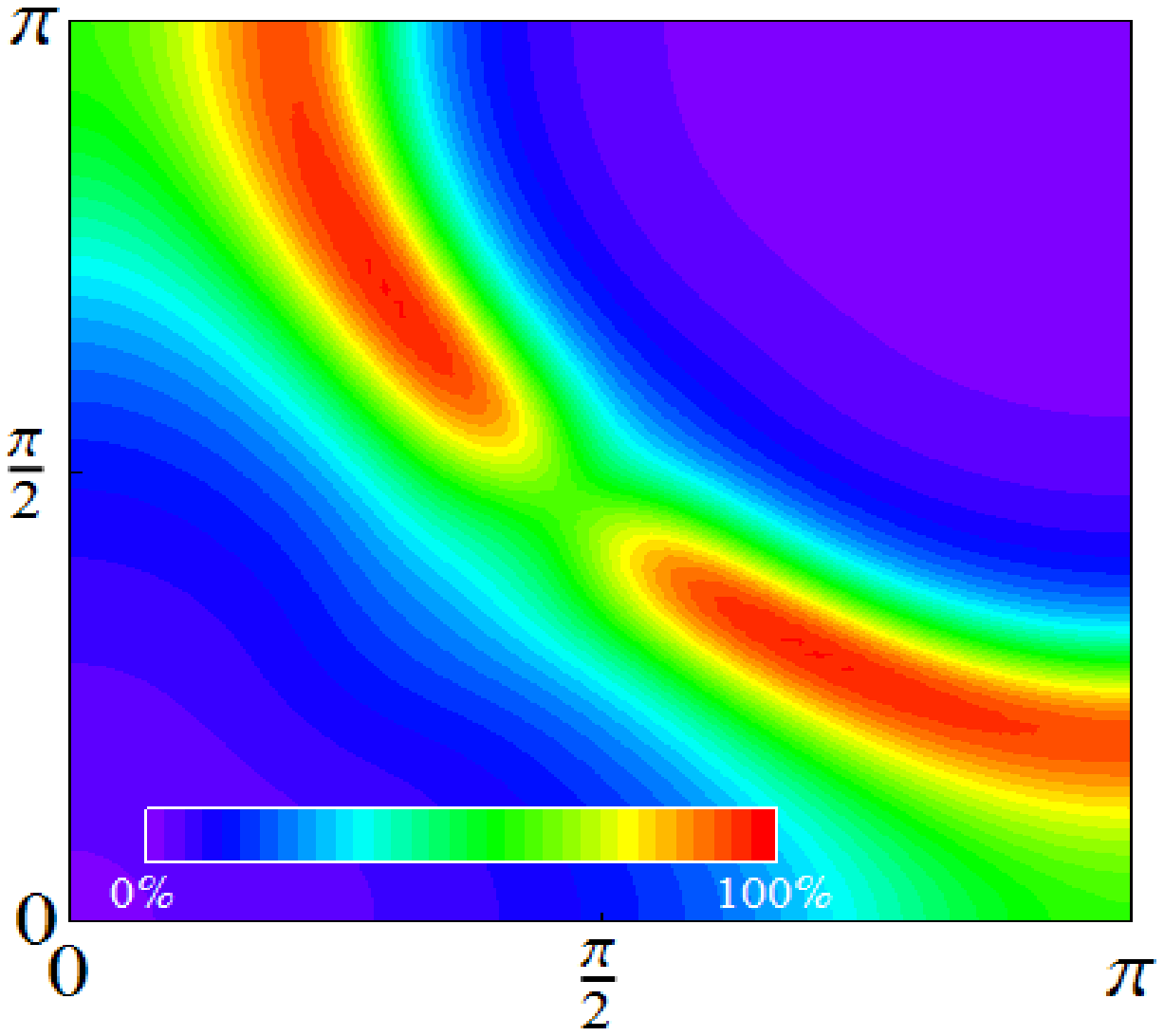}} %
\includegraphics[angle=-90,width=7cm]{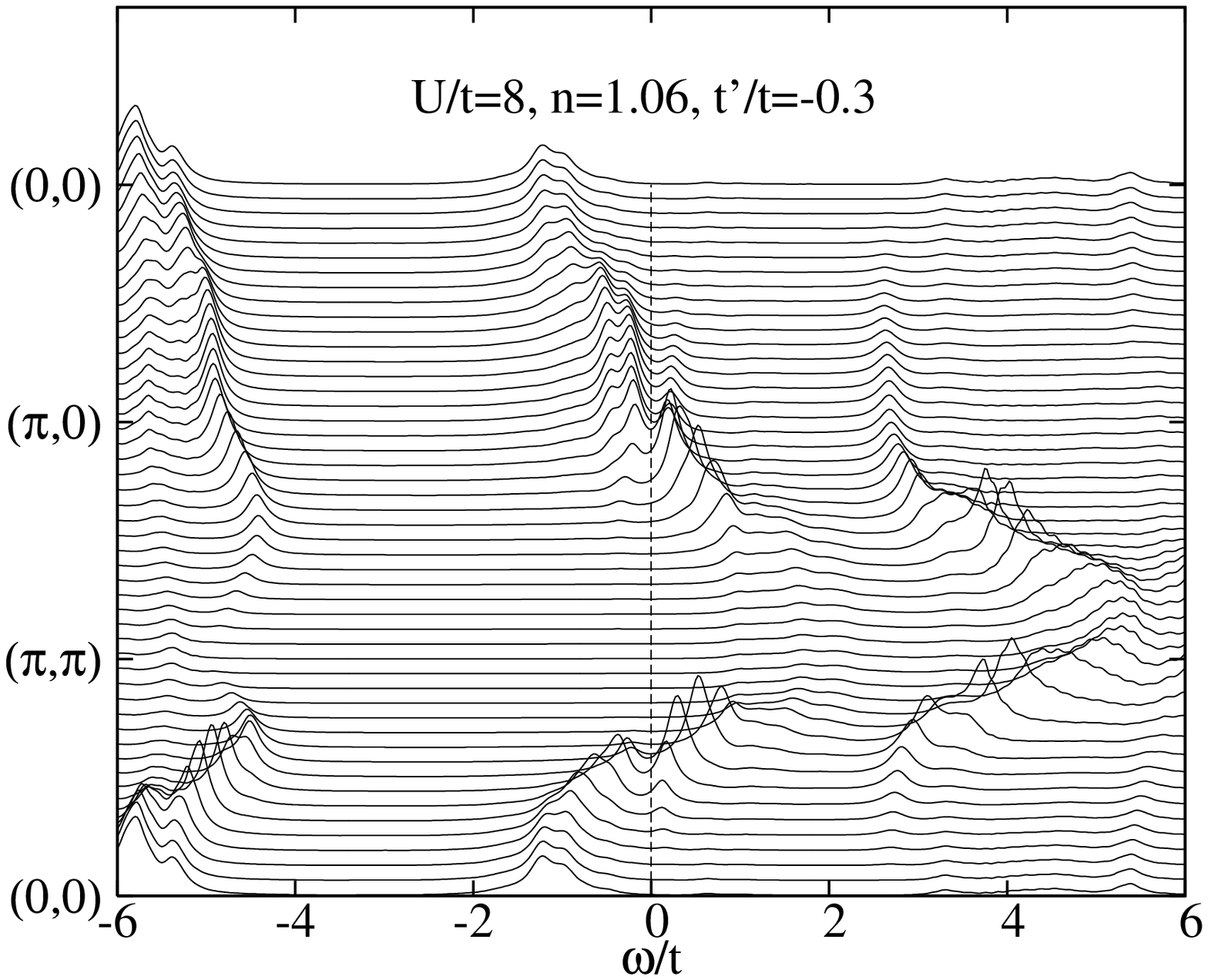}
\caption{(Color online) Same plots as in Fig. \protect\ref{spectra} but for
the electron-doped case $n=1.06.$ The maximum on the color scale corresponds to 104\% of the maximum of the data to stress finer details near the maximum. All other parameters are identical.}
\label{spectra_el}
\end{figure}

In Figs.~\ref{spectra} and \ref{spectra_el}, we present single-particle
spectra in the $d$-wave superconducting state for, respectively, hole doping $%
n=0.94$ and electron doping $n=1.06.$ The upper part of the figures
illustrates the momentum distribution curves (MDC) at the Fermi energy in
the first quadrant of the Brillouin zone, in other words a contour plot of
the single-particle spectral weight $A\left( \mathbf{k},\omega =0\right) $
with a finite energy resolution $\eta =0.1t$. The lower part contains energy
distribution curves (EDC) $A\left( \mathbf{k},\omega \right) $ as a function
of frequency in units of $t$ for various wave vectors $\mathbf{k}$ along
symmetry directions.

The MDC for hole doping, $n=0.94,$ in Fig. \ref{spectra} and electron
doping, $n=1.06$, in Fig. \ref{spectra_el}, are very similar to the
corresponding results in the normal state obtained earlier with the same
approach (Figs. 5 and 6 of Ref.%
~\onlinecite{KyungPseudogap:2006}%
, and Fig. 4 of Ref.%
~\onlinecite{civelli:2005}%
) and with CPT \cite{Senechal:2004, Hankevych:2006}, the latter having the
best momentum resolution. They display the striking asymmetry of the hole-
and electron-doped cases that can be measured by ARPES in the $d$-wave
superconducting phase \cite{Damascelli:2003} and in the pseudogap phase. More
specifically, for the hole-doped case the Fermi energy plot (MDC) shows
weight near $(\pi /2,\pi /2)$ with a clear suppression everywhere else. On
the other hand, the electron-doped case Fig. \ref{spectra_el} shows weight
in complementary areas of $(\pi ,0)$ and $(0,\pi )$ with the largest
suppression near $(\pi /2,\pi /2)$.

As shown in an earlier paper \cite{KyungPseudogap:2006}, it is likely that
for large values of $U$ the correlations responsible for the suppressed
weight in the normal state are of short-range antiferromagnetic origin,
although the precise nature of the magnetic fluctuations seems less
important than the fact that the system is close to a Mott insulator \cite%
{civelli:2005} and that there are short-range spatial magnetic correlations induced
by $J$. This is the strong-coupling mechanism for pseudogap discussed in
Refs.~\onlinecite{Senechal:2004} and \onlinecite{Hankevych:2006}. By contrast, the weak-coupling
mechanism does not need to be close to a Mott insulator. It involves long
wave length fluctuations \cite{Vilk:1997,Dare:2004,Motoyama:2007}
that can be explained by the Two-Particle Self-Consistent approach. The
latter mechanism seems to explain~\cite{Motoyama:2007} experimental results in slightly
underdoped Nd$_{2-x}$Ce$_{x}$CuO$_{4}$. For a discussion
of the pseudogap at large and small $U$ see Refs.
\onlinecite{Senechal:2004,Hankevych:2006}%
.


In the EDC for the hole-doped case on the lower part of Fig. \ref{spectra},
one can see in the $\left( 0,0\right) $ to $\left( \pi ,\pi \right) $
direction, a gap-like feature between $-t$ and $-2t$ that looks like the
so-called waterfall feature observed in ARPES experiments \cite%
{Xie:2007,Graf:2007,Graf:2006,Pan:2006,Meevasana:2007,Valla:2007}. This has some
similarity to what has been obtained in the $t-J$ model \cite{Tan:2007} but one
should be careful since at these energy scales it is likely that the
one-band Hubbard model differs from more realistic multi-band models \cite%
{Macridin:2005}. The gap-like feature here occurs between the spin-fluctuation induced band near the Fermi energy \cite{KyungPseudogap:2006}
and the rest of the lower-Hubbard band. There are differences with DMFT calculations since these are appropriate only for higher dimension~\cite{Vollhardt:2007}.

In the hole doped case of Fig. \ref{spectra}, we do not have the resolution
to address the question of whether there are two gaps, one associated with
superconductivity, and the other one associated with the pseudogap, as
discussed recently in the literature \cite{LeTacon:2006, Tanaka:2006,
AichhornGaps:2007,Civelli:2007}. This question is clearly most difficult to address
in the hole-doped compounds where the pseudogap and superconducting gap
occur in the same region of the Brillouin zone. In the electron-doped case
of Fig. \ref{spectra_el}, the position of the pseudogap or of the
antiferromagnetic gap near $(\pi /2,\pi /2)$ differs from the location of
the maximum superconducting gap. Hence, an earlier VCA study \cite%
{Senechal:2005} with better momentum resolution allowed one to see the
superconducting gap in the electron-doped case near $\left( \pi ,0\right) $
at the same time as the (pseudo)gap near $(\pi /2,\pi /2)$. In our case
also, the EDC in the lower part of Fig. \ref{spectra_el} does have both the
normal state gap near $\left( \pi /2,\pi /2\right) $ and the $d$-wave
superconducting gap near $\left( \pi ,0\right) $. The MDC color scheme in
the upper part of the same figure is on a relative scale. Hence, despite the
fact that there is both a pseudogap and a superconducting gap, the MDC
emphasizes that there is less weight in the pseudogap region near $\left(
\pi /2,\pi /2\right) $ than near the superconducting gap region near $\left(
\pi ,0\right) ,\left( 0,\pi \right) .$ Nevertheless, close inspection
reveals that regions of constant intensity carry less weight near $\left(
\pi ,0\right) $ than the corresponding normal state results in Ref.\
\onlinecite{KyungPseudogap:2006}%
. Also, the weight increases as one moves slightly away from the zone edge
along the Fermi surface, reflecting the superconducting gap. Experimentally,
the superconducting gap in the electron-doped compounds has been seen
independently of the pseudogap \cite{ArmitageGap:2001}.

\section{Conclusion}

\subsection{Summary}

We have shown that many of the non-BCS features exhibited by
superconductivity in the underdoped cuprates can be reproduced by CDMFT
calculations for the two-dimensional Hubbard model. For example, the fact that
the zero-temperature $d$-wave superconducting order parameter decreases as we move towards half-filling while the gap
in the single-particle density of states increases.
We have demonstrated semi-quantitative agreement with other cluster methods
and with experimental results for other quantities as well, such as the
ARPES spectrum and the doping range where superconductivity is stable.

We have also calculated the phase diagram that describes the competition
between antiferromagnetism and superconductivitity for $t^{\prime }=-0.3t$
for both hole and electron doping at various values of $U$. There is in
general homogeneous coexistence (superconducting antiferromagnetism or equivalently antiferromagnetic superconductivity) between
antiferromagnetism and $d$-wave superconductivity in the underdoped region.
This feature is quite generally observed in quantum-cluster methods,
variational approaches and mean-field theories that do not allow for
spin or charge density modulations on large length scales. In experiment, such coexistence seems to be the
exception rather than the norm. At large $U,$ the transition between the
coexistence phase and the pure $d$-wave state is first order.

Without attempting any precise fits to experimental data, it seems that the
phase diagram obtained for $t^{\prime }=-0.3t$ at intermediate coupling $U=8t
$ corresponds with the experimental phase diagram better than either smaller
$\left( U=6t\right) $ or larger $\left( U=12t\right) $ values of $U$.
Interestingly, adding the third neighbor hopping $t^{\prime \prime }=0.2t$,
as suggested by band structure, leads to an improvement of the qualitative
agreement between the calculated and observed phase diagrams, namely a
reduced tendency to coexistence and a larger electron-hole asymmetry. The
sensitivity to band parameters, expected at intermediate coupling where both
kinetic and potential contributions are comparable, hints towards
quantitative deviations from the universal $T_{c}\left( \delta \right)/T_{c}^{max}$
curve for compounds with different band parameters, such as La$_{2-x}$Sr$_{x}
$CuO$_{4}$ (LSCO) and YBa$_{2}$Cu$_{3}$O$_{7-x}$ (YBCO).

\subsection{Discussion}

The major questions left open by the present work are: a) whether the
Hubbard model by itself, without additional interactions, additional bands,
or without extrinsic disorder, can lead to a closer agreement with the experimental
phase diagram when additional broken symmetries are
allowed in the space of solutions \cite{Tranquada:2007, Kivelson:2007, bourges, KivelsonRMP:2003} b) if the additional broken
symmetries that are present in cuprates appear as secondary instabilities or
are essential to the physics of high temperature superconductivity.

Nevertheless, the Hubbard model seems sufficient to lead to the overall
phase diagram of the high-temperature superconductors and to explain many of
the observed non-BCS features, even in the absence of a final answer to the
last questions raised. In BCS theory, it is useful to think about the
mechanism for superconductivity, \textit{i.e.}\textbf{\ }what perturbation
makes the normal state unstable towards a superconducting ground state.
Similarly, in the weak-coupling Hubbard model, finite temperature DCA \cite%
{Maier:2007a,Maier:2007b,Maier:2007c} and Two-Particle Self-Consistent calculations (TPSC) \cite%
{Kyung:2003}, suggest that in this case it makes sense to think of $d$-wave
superconductivity as being mediated by antiferromagnetic fluctuations \cite%
{Bourbonnais:1986,Miyake:1986,Scalapino:1986,Carbotte:1999}%
, a generalization of the Kohn-Luttinger mechanism. In this weak- to intermediate-coupling
case, pairing occurs when the antiferromagnetic correlation length is large
\cite{Kyung:2003,Hassan:2007}. At strong coupling though, we have shown, in
agreement with other approaches, that the order parameter scales with the
superexchange interaction $J$, even when we do not allow for large
antiferromagnetic correlation lengths (which in the cluster treatements are
mimmicked by antiferromagnetic long range order). This shows that at strong
coupling, $d$-wave superconductivity can occur directly from the
superexchange~\cite{Kotliar:1988,Andrei:1988} $J$, without the need for an
antiferromagnetic ``glue'' or antiferromagnon exchange, as advocated by
Anderson \cite{Anderson:2007}. Still, at large $U$, the frequency dependence
of the order parameter has a complicated frequency dependence with multiple
scales present \cite{Haule:2007}.

Antiferromagnetism and $d$-wave superconductivity are two possible phases of
the strong-coupling Hubbard model. The normal state found in CDMFT is
unstable to either phase or to a coexistence phase, depending on parameters
and on doping. The CDMFT normal state contains both types of fluctuations,
$d$-wave and antiferromagnetic. At strong coupling it does not appear
necessary to think that antiferromagnetic fluctuations lead to $d$-wave
superconductivity or that $d$-wave superconducting fluctuations lead to
antiferromagnetism. This is perhaps best illustrated by the layered organic
conductors that can be modeled by a one-band Hubbard model on the anisotropic triangular lattice at half-filling. There, one finds, theoretically and experimentally,
a first order transition between antiferromagnetism and $d$-wave
superconductivity \cite{Sahebsara:2006,KyungBEDT:2006}, so both phases are
different instabilities of the same Hamiltonian.

CDMFT and a number of methods show that maximum pairing (maximum value of
the $d$-wave order parameter) occurs at intermediate coupling, a range
appropriate for the high-temperature superconductors. So both weak and
strong-coupling features may appear \cite{Maier:2007a,Maier:2007b,Maier:2007c}.

One of the ways to distinguish weak and strong coupling limits is by the
behavior of the $d$-wave order parameter as one approaches half-filling. At
strong coupling, it decreases as $n\rightarrow 1$ when $U$ is larger than
the critical $U$ for Mott localization. The phase fluctuations and
short-range antiferromagnetic correlations that lead to the decrease need
not be long range: They occur within the cluster. At weak coupling on the
other hand, the decrease in the order parameter near half-filling does not
occur unless antiferromagnetic fluctuations with correlation lengths large
enough to create the pseudogap are included \cite{Kyung:2003, Hassan:2007}.
Exactly at half-filling, weak and strong coupling limit are separated in the
normal state by the Mott transition \cite{Parcollet:2004, Nevidomskyy:2007},
in the absence of magnetic long range order. Away from half-filling, one can
postulate that when the correlation length of the antiferromagnetic
fluctuations become of the order of the lattice spacing, then $J$ by itself
suffices to lead to pairing at strong coupling. There may not be a sharp
phase transition between the two extreme limits.

Quantum cluster methods are powerful numerical ways of attacking the problem
of high-temperature superconductivity. They clearly show that many aspects
of the physics of the high temperature superconductors are contained in the
Hubbard model. The convergence between the results of these methods and
those of other recent numerical approaches suggests that we are closing in on
methods that can provide a quantitative solution of this problem.

\section*{Acknowledgements}

We thank P. Fournier and D.J. Scalapino for useful discussions. The present
work was supported by NSERC (Canada), FQRNT (Qu\'{e}bec), CFI (Canada),
CIFAR, the Tier I Canada Research chair Program (A.-M.S.T.) the NSF under
grant DMR 0528969 (G.K.) and MIUR PRIN Prot. 200522492 (M.C.). Computations
were carried out on the RUPC cluster at Rutgers, on the Elix2 200-cpu
Beowulf cluster and on the 868-cpu Dell cluster of the R\'{e}seau qu\'{e}b%
\'{e}cois de calcul de haute performance (RQCHP), both at Sherbrooke. G.K.
and A.-M.S.T. thank the Aspen Center for Physics where the final version of
this paper was drafted.




\end{document}